# Midlatitude Cirrus Clouds and Multiple Tropopauses from a 2002-2006 Climatology over the SIRTA Observatory

Vincent Noël [1] and Martial Haeffelin [1]

[1] *Institut Pierre-Simon Laplace / Laboratoire de Météorologie Dynamique, Ecole Polytechnique, Palaiseau, France*

Corresponding author :

Vincent Noël

Laboratoire de Météorologie Dynamique

Ecole Polytechnique

91128 Palaiseau

FRANCE

vincent.noel@lmd.polytechnique.fr

**Abstract**

This study present a comparison of lidar observations of midlatitude cirrus clouds over the SIRTA observatory between 2002 and 2006 with multiple tropopauses (MT) retrieved from radiosounding temperature profiles. The temporal variability of MT properties (frequency, thickness) are discussed. Results show a marked annual cycle, with MT frequency reaching its lowest point in May (~18% occurrence of MT) and slowly rising to more than 40% in DJF. The average thickness of the MT also follows an annual cycle, going from less than 1 km in spring to 1.5 km in late autumn. Comparison with lidar observations show that cirrus clouds show a preference for being located close below the 1$^{st}$ tropopause. When the cloud top is above the 1$^{st}$ tropopause (7% of observations), in 20% of cases the cloud base is above it as well, resulting in a cirrus cloud "sandwiched" between the two tropopauses. Compared to the general distribution of cirrus, cross-tropopause cirrus show a higher frequency of large optical depths, while inter-tropopause cirrus show almost exclusively low optical depths ($\tau < 0.03$ in 90% of cases) typical of subvisible clouds. Results suggest the occurrence of inter-tropopause cirrus clouds is correlated with the frequency of multiple tropopauses.

# I. Introduction

Clouds have been identified by the IPCC as one of the most important problems to tackle regarding climate prediction (IPCC 2001), as their radiative and dynamic influence on the Earth's climate is complex and not yet fully understood. Exercises in climate model comparison reveal divergences even in the sign of cloud feedback to CO2 doubling scenarios (e.g. Cess et al., 1990; Bony et al. 2004). The role of high altitude ice clouds, herein after referred to as cirrus clouds, is particularly difficult to characterize. As they are optically thin (i.e. semi-transparent), they often lie below the detection threshold of most passive remote sensing instruments such as those present on meteorological and research satellite systems. Recent studies suggest the existence of an entire population of subvisible ice clouds (Nee et al. 1998, Goldfarb 2001) whose radiative influence (Prabhakara et al. 1993) is totally overlooked in current radiative budget calculations. Additionally, the optical and microphysical properties of cirrus clouds are extremely variable (Cantrell and Heymsfield 2005). Hence, it is difficult to ascertain if their properties can be correctly parameterized at small or large scales, as they strongly depend on local phenomenonas - even when only considering latitude-constrained areas (Heymsfield and Miloshevich 2003). Since recent results suggest that cirrus cloud cover may be increasing due to human activities (Zerefos et al. 2003, Minnis et al. 2004), improving our knowledge of these clouds is a current and important issue.

The conditions in terms of thermodynamics, dynamics and ice forming nuclei are keyelements that influence the formation of high-altitude tropospheric clouds (Sassen 1992, Mace et al. 2001, Korolev et al 2003). Since cirrus clouds are found at high altitudes, their formation process and further evolution can be sensitive to atmospheric conditions in the Upper Tropos-phere and Lower Stratosphere (UTLS). Studying UTLS conditions could give hints on what drives cirrus cloud formation and how to better predict their properties. The tropopause temperature inversion has a constraining effect on cloud altitude, and several studies have focused on the properties of cirrus clouds near the tropopause, especially at tropical latitudes (McFarquhar et al. 2000; Peter et al. 2003; Garrett et al. 2005). Hence variability in the thermodynamic structure of the tropopause could affect the properties of cirrus clouds near the UTLS transition. One such structure is the occurrence of multiple inversions in the UTLS area resulting in the formation of Multiple Tropopauses (MT). More than just a theoretical concept, multiple tropopauses are symptoms of actual physical phenomena (Sect. III) that can help detect and analyze specific atmospheric conditions of cloud formation.

This paper presents a study of MT at midlatitudes, and investigates their possible influence on the occurrence, persistence and optical properties of high-altitude tropospheric clouds. The observation dataset and the retrieval techniques is presented in Sect. II. Statistical properties of the retrieved MT are presented in Sect. III and compared to existing litterature. The frequency of occurrence and properties of cirrus clouds in case of MT are then compared to the rest of the database in Sect. IV. Results are discussed and conclusion is given in Sect. V.

## II. Observations

To investigate a possible relationship between multiple tropopauses and ice clouds, the present study uses collocated in-situ and remote sensing observations made between 2002 and 2006. Tropopause levels were retrieved from temperature profiles, obtained through a dataset of radiosoundings launched twice daily (00 and 12 UT) from the Meteo France

Trappes regional center (48.8°N, 2.0°E), 20 km to the West of Paris. The dataset contains over 3000 radiosonde profiles gathered over the same period. The simultaneous study of ice clouds involved finding a data source suited to the detection and analysis of optically thin clouds.

During the past decades, numerous techniques have been developed to detect cloud occurrences from e.g. passive remote sensing observations by spaceborne radiometers (Rossow and Garder 1993). These methods are generally successful in estimating the spatial and temporal coverage of clouds with moderate to high optical depths τ. However, due to their limited optical thickness, ice clouds are usually difficult to detect via these conventional means. Especially in the lower end of optical thickness, subvisible cirrus (τ < 0.03) account for more than 20% of cirrus clouds (Goldfarb et al. 2001) but are in effect invisible to passive remote sensing, typically limited to optical depths above 0.1 (Stordal et al. 2005). In addition, several studies have shown that retrievals of high-altitude cloud properties using passive cloud imager data, such as cloud top height (e.g. Naud et al. 2004; Daloze and Haeffelin 2005) or microphysical properties (e.g. Chiriaco et al. 2004), can include significant biases due to the semi-transparent nature of those clouds. This precludes the use of passive remote sensing data to study interactions between the thermodynamic structure near the UTLS and cirrus cloud properties. Other observations are therefore required to obtain a realistic estimate of the ice cloud cover, and lidar measurements have become the *de facto* standard when considering this particular problem, thanks to their high sensitivity to optically thin atmospheric layers. Moreover, most modern lidar systems are equipped with the ability to read the polarization state of the backscattered light (Sassen 1991), providing insights into the cloud phase (Wang and Sassen 2001) and microphysical properties (e.g. ice crystal shapes, Noel et al. 2004).

In the present paper, lidar observations are provided by the 532 nm LNA (*Lidar Nuages Aerosols*) lidar system, located at the SIRTA Observatory (*Site Instrumental de Recherche par Télédétection Atmosphérique*, 48.71°N 2.21°E, Haeffelin et al. 2005), 15 km east of Trappes. This lidar features a narrow field-of-view telescope (0.5 mrad) dedicated to the observation of high-altitude tropospheric clouds, and is equipped to retrieve 20 vertical profiles of backscattered light and depolarization ratio $\delta$ by second, with a 15 meters resolution. The first lidar observations stored in the SIRTA database date back to 2002 (Matthias et al. 2004), and routine operation (5 days a week) was initiated in 2004, totaling today more than 500 days of observations. The STRAT algorithm (Morille et al. 2006) was applied 1) on the backscattered light profiles to separate cloud layers from the molecular background (using wavelet transform analysis), and 2) on the depolarization ratio to unequivocally flag ice clouds (only clouds producing values of $\delta > 0.2$ were considered). Optical depths were then retrieved for ice cloud layers using a standard transmission-loss algorithm (e.g. Platt, 1973; Platt et al. 1998). This algorithm derives the extinction produced by a given cloud layer (hence, its optical depth) by comparing the molecular backscatter in the free troposphere above and below the layer. This requires a significant signal-to-noise ratio above the given layer - i.e. only layers that were penetrated all the way through by the lidar were considered. This criteria means that a lot of cloud layers were discarded in the process; it is assumed that the volume of observation is still sufficient to paint an accurate representation of cirrus clouds.

# III. Multiple Tropopauses over the SIRTA

Despite its central role in atmospheric sciences, the tropopause can still behave in unusual, surprising ways. For instance, depressions at midlatitudes (e.g. on the Eastern U.S. coast) can drag cold air masses from the arctic, resulting on occasion in tropopauses as low as 6 km. In the Tropics, the separation between the troposphere and the stratosphere is often defined as the point of temperature minimum, but in midlatitude regions the thermal structure of the tropopause layer can be quite complex. The thermal or lapse-rate tropopause definition is based on the variability in lapse rate (-dT/dZ) in an atmospheric temperature profile. In the most commonly used definition, the conventional tropopause is defined by the World Meteorological Organization (WMO) as the "lowest level at which the lapse rate decreases to 2°C/km or less, provided also the averaged lapse rate between this level and all higher levels within 2 km does not exceed 2°C/km" (WMO 1957). However, the WMO definition also allows additional tropopauses above the first one, if the average lapse rate between any level above the first tropopause and all higher levels within 1 km exceeds 3°C/km. This technical definition, which is used in the rest of the present study, actually reflects dynamical disturbances to the temperature profile (e.g. jet streams or upper level fronts, Nadstrom et al. 1989) resulting in multiple temperature inversions in the UTLS (Shapiro 1980), that can lead to tropopause foldings (Sprenger et al, 2003) and mixing of stratospheric and tropospheric air (Hoinka 1998). According to a recent climatology of MT using Global Positioning System radio occultation data between 2001 and 2004 (Schmidt et al. 2006), MT frequency and thickness (defined as the altitude difference between the lowest and the highest tropopauses) exhibit a strong dependence on latitude and season. Their results show the occurrence of MT is almost null near the equator (20°S-20°N), while it varies between 80% (DJF) and 45% (JJA) at 40°N. The seasonal dependency is less intense in the southern hemisphere, but occurrences of ~70% are still reached in the southern hemisphere winter (JJA). MT thickness is correlated to MT occurrence frequency, reaching maxima of 5.2 km at 40°N (DJF) and 4.0 km at 40°S (JJA).

Analysis of temperature profiles from the Trappes radiosoundings (49°N) on the 2002-2006 period shows that Multiple Tropopauses occur in 31.4% of cases, with a $3^{rd}$ tropopause in 10.1% of cases and a $4^{th}$ in 4.3% of cases (additional tropopauses were considered to be negligible). Figure 1 shows the annual cycle of MT occurrences, averaged over the 2002-2006 period. The vertical bars represent the 2002-2006 interannual variability (one standard deviation), confirming the significance of the annual cycle. The tropopause temperature and pressure are known to exhibit quasi-biennial oscillation and long-term fluctuations (Angell and Korshover 1973) than can affect both the annual cycle and the interannual variability. The annual MT occurrence cycle begins with high occurrences (~35%) in January, followed by a steep decline during spring (reaching a ~18% minimum in May), then finally rising slowly during summer and autumn to reach again high occurrences (~43%) in December. High MT occurrences in winter (DJF) coincide with more frequent occurrences of fronts. These values are consistent with results from Schmidt et al. 2006, who found occurrences over France in the 45-55% range in winter (DJF) and 20-35% range in summer (JJA). Figure 2 shows the annual cycle of tropopause altitudes and the mean Inter-Tropopause Thickness (ITT, i.e. the distance between the lowest and highest tropopauses) based on 2002-2006 data. The minimum and maximum altitudes are reached in January-April and June-September respectively. The monthly mean altitude of the first tropopause varies 1.6 km between winter and summer (from 10.2 to 11.8 km), while that of the second tropopause varies almost 2.7 km (from 10.25 to 12.9 km). The annual variations of additional tropopauses (3$^{rd}$ and 4$^{th}$) exceed 3 km, with the maximum variation showed by the 3rd tropopause: from 10.75 km in April to more than 14 km in August.

The Inter-Tropopause Thickness shows a similar annual evolution (thick curve in Fig. 2), from a 900 m minimum in May to a 1500 m maximum in October-November. These values are slightly lower than the averages found in the Schmidt et al. (2006) study over France (1.1-1.5 km for JJA and 2.2-2.8 km for DJF), but the general trend is consistent. The ITT increase between May and October is correlated with the increase in multi-tropopause frequency over the same period (Fig. 1), however the sharp drop in ITT between November and December is not found in multi-tropopause frequency (where the decrease happens slowly between December and May). It should also be noted that in Fig. 2 the ITT should not necessarily follow distance between the 1$^{st}$ and 4$^{th}$ tropopauses - for its calculation the highest tropopause is used on a case-by-case basis, and a 4$^{th}$ Tropopause was only present in 4.3% of profiles. Partly because of this, the ITT is not directly correlated with the tropopause altitudes: for instance the period between June and September is marked with stable high tropopauses, but meanwhile the ITT keeps increasing. This can be explained by an increase in the absolute number of tropopauses during the same period. The ITT reaches its peak in October as the MT frequency reaches stability, and begins decreasing simultaneously with tropopause altitudes. The average ITT is constrained between 0.6 and 1.5 km (Fig. 2, more than 50% of all observations), however the individual values spread out over a very large dis-tribution following a gamma shape with a major mode centered on 750 m. The absolute minimum for an ITT is 215 meters, but values larger than 3 or 4 km are not uncommon.

## IV. Ice Clouds and Multiple Tropopauses

The SIRTA Cirrus cloud 2002-2004 dataset (322 cases) has been described extensively in Noël et al. 2006, including a presentation of cirrus occurrence statistics. After a review of the 2005 observations, it appears the conclusions reached in this paper still hold true for the extended dataset. The number of cirrus sightings increases smoothly with decreasing temperature between -10°C and -45°C, to reach its maximum in the (-55°C, -45°C) temperature range, followed by a quick drop to zero occurrence at -75°C and colder. More than half of the cirrus clouds are observed in areas where the relative humidity ranges between 80 and 120%. The frequency of cirrus detection is variable over the course of each year, but it is not possible to ascertain if this is due to an actual change in the population of cirrus clouds or to sampling effects (Protat, 2006).

Because the lidar is not operating continuously, and cirrus are sometimes hidden from it by lower clouds, the frequency of cirrus sightings (from lidar) compared to the frequency of multiple tropopauses (from radiosoundings) is a pointless exercise which will not be attempted here. On the other hand, comparing the properties of multiple tropopauses and cirrus clouds on days when both are detected and fully visible could expose correlations between the two phenomena.

## IV. 1. Cirrus Tops and Bases

This study focuses on the relationship between the tropopause(s) and cirrus clouds, thus only clouds whose top altitudes are located higher than 7 km will be considered - this amounts to 198794 lidar profiles, or 1656 hours of cirrus observation. The distance between cloud top altitudes and the first (lowest) tropopause is shown in Fig. 3a for all clouds with top altitudes above 7 km. Negative (resp. positive) values indicate cloud top altitudes below (resp. above) the 1$^{st}$ Tropopause. Clearly most cirrus clouds are entirely located below the 1$^{st}$ tropopause. For such clouds, we find no impact of the occurrence of single or multiple tropopauses on the distribution of cloud top and base altitudes. Not surprizingly, the distribution (Fig. 3a) reveals a large frequency of occurrence of cloud top altitudes close to the tropopause - i.e. cirrus clouds are very likely stuck right under the tropopause.

Cloud with tops above the first tropopause occur in more than 5000 profiles (2.5% of the observations, 40 hours of operating time), out of which a strong 86% happen in a multi-tropopause situation. For those clouds, a study of distance between cloud top altitude and the 2nd tropopause should naturally come next - however, after review this distance does not yield very meaningful results, as it is itself strongly dependent on the ITT which varies widely on a scale of several kilometers, i.e. in the same range as cloud top altitudes (Fig. 3a). Instead, we use the ratio of the distance between cloud top and first tropopause to the distance between the 1$^{st}$ and 2$^{nd}$ tropopauses ( $\frac{z_{Cloud\,Top} - z_{Tropo1}}{z_{Tropo2} - z_{Tropo1}}$ ) to evaluate how much of this region is filled by clouds. This ratio is 0 (resp. 1) for cloud tops stuck to the 1$^{st}$ (resp. 2$^{nd}$) tropopause. Its distribution (Fig. 3b) is found to be bimodal and in 50% of cases the ratio is below 0.3, meaning the related clouds occupy a limited region between the 1st and 2nd tropopauses, with their top in the lower 30% of the inter-tropopause zone. However, in 33% of cases the ratio is above 0.7, meaning cloud tops are in the higher 30% of the same zone (i.e. these clouds span the entire region). In only 16% of cases was the ratio between 0.3 and 0.7. Almost no cirrus cloud was detected with a top altitude above the 2nd tropopause. Therefore, it appears that clouds that cross the lowest tropopause live in an unstable temperature profile, which leads a large fraction of them to rise or expand vertically until they reach the next tropopause.

Next we study the distribution of cloud base altitude for clouds that extend above the first tropopause. Fig. 4 shows that tropopause-crossing cirrus clouds have most frequently their base altitude less than 2 km below the 1st tropopause, with a maximum frequency of occurrence between -1.5 and -1 km. The distribution of cloud base altitudes is wide, with some tropopause-crossing cloud bases as low as 4 km below the 1st tropopause. The cloud base altitude distribution is much narrower for clouds that are entirely contained in the inter-tropopause zone (positive values in the Fig. 4 histogram), with almost all cloud base altitudes within 1 km above the 1st tropopause. These clouds are, in effect, "sandwiched" between the two lowest tropopauses. Only a very small, negligible number of clouds (not shown) have their bases above the second tropopause.

By considering Fig. 3 and 4, it is possible to classify cirrus clouds in 4 categories (Table 1) defined by the position of their boundaries with respect to the tropopause(s): (1) tropospheric cirrus, with cloud top altitude more than 500 m below the 1st tropopause (83 %); (2) "ceiling" cirrus, with cloud top altitude less than 500 m below the 1st tropopause (10 %); (3) cross-tropopause (CT) cirrus (5%) and (4) inter-tropopause (IT) "sandwich" cirrus contained in the inter-tropopause zone (2%). As this paper focuses on the relationships between multiple tropopauses and cirrus clouds, the following sections are devoted to the last two categories, trying to highlight their specific properties compared to the rest of the cirrus population. Even if, comparatively, they represent only a small part of the entire SIRTA dataset, the absolute population of these clouds is significant enough - for instance, 16 cases of inter-tropopause clouds were identified for the single year 2005. As an example, a cloud observed on March 17$^{th}$ 2005 is shown in Fig. 5a. The first and second tropopauses are marked with dashed lines, at 11.5 and 12.5 km respectively, and are clearly apparent on the relevant temperature profile from the 12 UT radiosounding (Fig. 5b). Note that temperature inversion of the first tropopause is significantly weaker than that of the second tropopause. During the 3.5 hours of observation, the cloud remains inside the limits defined by the 1$^{st}$ and 2$^{nd}$ Tropopauses. The 2$^{nd}$ Tropopause acts as a cloud top ceiling.

### IV. 2. Geometrical and Optical Thickness

The distribution of cloud geometrical thickness for CT and IT cirrus clouds (Fig. 6) is multi-modal, with a major mode centered on 0.5 - 1 km, and a minor mode centered on 3 - 3.5 km. Study showed that these two modes cannot be independently attributed to CT and IT clouds; however they do not appear when considering all cirrus clouds, and therefore could be related to the tropopause behavior and properties. However, relationships between these modes and the distance between cloud and tropopause altitude, or the inter-tropopause thickness could not be established. These two modes are in fact related to seasonal variations: clouds observed between May and November (dark) have a distribution strongly biased towards low values of geometrical thickness (0 - 1.5 km) - i.e. they are thinner; clouds observed between December and April (light) equally include thin clouds (with a similar mode centered on small values), but a new mode appears between 2 and 4 km, centered on 3.25 km. This mode is totally absent of the May to November observations; thus it follows that geometrically thick cirrus clouds close to the tropopause are only observed between December and April, i.e. a period marked by a high frequency of multiple tropopauses (DJF, Fig. 1) and a simultaneous decrease in the inter-tropopause thickness (Fig. 2). This is consistent with the observation that tropopause-crossing cirrus clouds are most frequently observed in conjunction with multiple tropopause conditions (Sect. 4.1).

Next we study the optical depth distributions for the four different cirrus cloud classes defined in Sect. 4.1. Figure 7 represents the cumulative density functions of optical depth $\tau$ for all cirrus clouds (thick green line) and for the four different cirrus clouds classes (Sect. IV 1). In this graphical representation, a steep increase signals a high density of the related optical depth, and conversely, low slopes and horizontal sections convey low density. Thus the absence of any extreme slope in the all-cirrus curve shows optical depths are generally evenly distributed between $10^{-3}$ and 1, with the most frequent being found between 0.05 and 0.2. The distributions are analyzed in terms of three optical depths classes, namely thick ($\tau > 0.3$), semi-transparent ($0.03 < \tau < 0.3$) and subvisible ($\tau < 0.03$) using the definition provided by Sassen et al. (1989). The all-cirrus distribution reveals 40% subvisible, 50% semi-transparent and 10% thick cirrus clouds. Free-tropopause and "ceiling" cirrus (i.e. clouds with tops below the first tropopause, Sect. IV.1), shown in black and red lines, follow very closely the general distribution, which is consistent considering they represent respectively 83% and 10% of all cirrus clouds. CT cirrus (blue line) are found to be optically thicker than the general distribution, with 25% subvisible, 65% semi-transparent and 10% thick cirrus clouds. The horizontal section for $\tau < 10^{-3}$ on the blue curve signals the nearly total absence of very low optical de-pths that exist, although in small quantities, in the general dataset (green line) ; on the other hand, the steep slope of the blue curve around $\tau \sim 1$ suggests the distribution peaks around these values, while the general dataset is more evenly distributed. IT cirrus clouds (purple line) have evidently far lower optical depths than the other groups, with 75% subvisible, 25% semi-transparent and virtually no thick cirrus clouds. Half of the clouds in that group show optical depths below 0.007. The purple curve is rugged compared to the others, due to the smaller number of observations in that dataset.

Finally we study the relationship between the optical and geometrical thickness of the different cloud classes. The density of cirrus clouds as a function of optical depth $\tau$ and geometrical thickness $h$ (Fig. 8) shows a high concentration of optical depths between 0.005 and 0.02 (consistent with Fig. 7) but in addition highlights a high correlation between the two properties. Low optical depths ($\tau < 10^{-3}$) are exclusively found for clouds thinner than 0.5 km, but as optical depths increase the range of possible geometrical thickness rises as well. Subvisible cirrus clouds ($\tau \leq 0.03$) can bear any geometrical thickness between 0 and 2.5 km, with highest frequencies for $0.2 < h < 0.8$ km ; the widest range of cloud thickness ($0 < h < 4$ km) is found for $\tau \sim 0.3$. For $\tau > 0.3$ a minimum geometrical thickness appears, which increases with optical depth: for instance, cirrus clouds with $\tau \sim 1$ are always thicker than 0.5 km. When considering densities from the other axis, very thin clouds ($h < 0.5$ km) can bear any optical depth between $10^{-4}$ and 0.3 (with a maximum frequency for $0.001 < \tau < 0.03$, i.e. subvisible cirrus). The possible minimum optical depth then gradually increases from $10^{-4}$ to 0.05 for cloud geometrical thickness between 0.5 and 4 km. No subvisible cirrus cloud is thicker than 2 km. The relationship between optical and geometrical thickness is consistent for all clouds considered in the present study. Dots in Fig 8 show observations relative to CT and IT cirrus clouds, they follow the general distribution quite well, with a relatively higher dispersion of optical depths for cloud thickness above 3 km that might just be due to the relative sparseness of data.

## V. Discussion and Conclusion

In this study, multiple tropopause altitudes were retrieved over the SIRTA observatory from radiosoundings (Sect. III), producing results that are consistent with a previous GPS study (Schmidt et al 2006). Using lidar observations it was possible to compare these retrievals with cirrus cloud top and base altitudes (Sect. IV.1). Distributions of the distance between cloud boundaries and the tropopause layer were presented, and used to classify the clouds in 4 categories. Clouds with tops above the first tropopause show distinctively different optical depths distributions (Sect. IV. 2, Fig. 7), but the optical-to-geometrical thickness relationship is not affected (Fig. 8). A study of the distribution of wind speed and direction showed similar distributions for all 4 categories of cirrus clouds (mostly winds coming from the Atlantic Ocean, see Noël et al. 2006), with no specific values for CT and IT clouds - hence, related figures were not presented.

The study of cloud top and base altitudes with respect to tropopause altitudes yields several conclusions. Firstly, most cirrus clouds are totally contained below the first tropopause (93%), but a non-negligible part either cross the tropopause (5%) or is totally contained between the first and second tropopause (2%). For the first group, the presence of multiple tropopauses has close to no effect on their altitude and thickness. However, the frequency of occurrence of the clouds increases with decreasing distance between the cloud top and the first tropopause - with almost 30% of cloud tops being located less than 2 km below the first tropopause (Fig. 3a). In a similar fashion, clouds become more frequent as cloud bases get closer to the tropopause, however the maximum frequency is found 0.2-1.5 km below the tropopause. This is consistent with cirrus clouds stuck the tropopause considering this is also the most frequent cloud thickness range (Fig. 8). For CT and IT clouds, the effect is much stronger: tops are either very close to the first tropopause, or stuck all the way to the second tropopause (Fig. 3b), while IT cloud bases are all within 0.5 km above the first tropopause (Fig. 4).

Cloud categories are even more distinct when considering optical depth distributions (Fig. 7). While clouds located entirely below the first tropopause are very similar to the general distribution (they mostly define it), clouds that cross or are above the tropopause differ significantly: CT clouds have a very low density of subvisible clouds ($\tau < 0.03$) and a large concentration of semi-transparent ($\tau > 0.03$) and thick ($\tau > 0.3$) clouds, while the picture is reversed for IT clouds which are predominantly in the subvisible category and present no thick cloud. A simple explanation for the relatively high optical depth of CT clouds is that by definition, CT clouds can extend further up than BT clouds. Since the optical-to-geometrical thickness relationship is not affected by the cloud location (Fig. 8), it follows that CT clouds should be optically thicker than BT clouds. Regarding the cloud origin, the simplest hypothesis is that CT clouds originate from within the free troposphere, where they have room to expand, shrink and move vertically (either upwards or downwards), crossing the first tropopause at some point during their life cycle, in particular when multiple tropopauses occur resulting in a weak thermal inversion at the first tropopause level. Looking at IT clouds, their low optical depths cannot be solely explained by their low geometrical extent (constrained by the double tropopauses): Fig. 8 shows that even thin cirrus layers (i.e. less than 1 km thick) can adopt a wide spectrum of optical depths, up to 1. Other explanations will have to be investigated. Regarding their origin, the simplest hypothesis is that IT clouds originate from within the inter-tropopause zone, where they are constrained by the double temperature inversion (Fig. 3 and 4) and thus keep a very low optical depth. 82% of IT and CT clouds with a ratio of

$$\frac{z_{Cloud\ Top} - z_{Tropo}}{ITT} > 0.5$$ (Fig. 3b) have their base within 0.5 km of the first tropopause - in other words, clouds that extend up to the second tropopause have, in most cases, their bases stuck to the first tropopause. This suggests that IT clouds very rarely cross the tropopause to extend downward in the troposphere, confirming the theory that CT clouds originated in the free troposphere and later extended upwards in the inter-tropopause zone. Of course, these are simple considerations given the actual observations; cirrus generation mechanisms involves complex processes and further studies will be required to reliably assess the origin and formation of each cirrus category.

Considering these results, it seems at first glance that cirrus clouds created *in situ* in the inter-tropopause zone develop to become almost exclusively subvisible cirrus, and never extend down to develop into optically thicker cloud systems: their development seems hindered by the two temperature inversions found in multi-tropopause situations. Since IT clouds are so high and cold, they have a significant impact in terms of climate change due to their influence on the net radiation budget at the top of the atmosphere (Brown and Francis, 1995). These clouds seems quite rare, at least over the SIRTA (2% of all cirrus occurrences), but since their formation seems correlated with multiple tropopause events, their frequency could in-crease at other latitudes that show much higher occurrences of multiple tropopauses (e.g. up to 80% in the Mediterranean region or the United States, Schmidt et al. 2006).

Finally, it is important to remember that ground-based lidars alone cannot provide a exhaustive coverage of cirrus clouds near the tropopause, as (1) some of the cirrus clouds cannot be penetrated all the way through due to their high optical depth, and were not considered in the present study; and (2) an unknown fraction of cirrus clouds is hidden by optically thick, low-level clouds. This shouldn't modify the optical depth distribution found for cirrus categories in the present study (as cirrus clouds have generally low opacity), however an unknown quantity of cirrus clouds were not present in the observations, and could sway the importance of a given category one way or another. A possible way to study the entire population of cirrus clouds would be to use combined lidar and radar profiles from spaceborne instruments, i.e. viewing the atmosphere from above. Such observations will soon be available from the recently launched CALIPSO (Winker et al. 2003) and CloudSat (Stephens et al. 2002) spaceborne platforms. In addition to the use of these observations, future work include (1) considering a definition of tropopause based on potential vorticity, which might allow a better understanding of the dynamic features associated with multiple tropopauses and the occurrence of cirrus clouds, and (2) investigating if cirrus in each category share a common origin and formation mechanisms, by correlating back-trajectories of air masses in observed clouds with spatial imagery and models.

# VI. References

- Angell J. K. and J. Korshover 1974: Quasi-biennial and long-term fluctuations in tropopause pressure and temperature, and the relation to stratospheric water vapor content. *Mon. Wea. Rev.* **102** 29-34.

- Bony, S. et al, 2004 : On dynamic and thermodynamic components of cloud change. *Climate Dynamics*. DOI: 10.1007/s00382-003-0369-6, Vol. 22, No. 2-3, pp.71-86

- Brown P. R. A. and N. Francis, 1995: Improved measurements of the ice water content in cirrus using a total water probe. *J. Atmos. Oceanic. Tech.* **12**, 410-414.

- Cantrell W. and A. Heymsfield 2005: Production of Ice in Tropospheric Clouds: A Review. *Bull. Am. Met. Soc.* **62** (7) 2352-2372.

- Cess, R. et al, 1990 : Intercomparison and interpretation of climate feedback processes in 19 atmospheric general circulation models. *J. Geophys. Res.*, 95, 16601-16615

- Chiriaco M., H. Chepfer, V. Noel, A. Delaval, M. Haeffelin, P. Dubuisson, P. Yang 2004: Improving Retrievals of Cirrus Cloud Particle Size Coupling Lidar and Three-Channel Radiometric Techniques. *Mon. Wea. Rev.* **132** 1684-1700.

- Daloze J-F., M. Haeffelin, Validation of SAFNWC/MSG Cloud Top Height using ground-based lidar and radar measurements. Visiting Scientist report, CMS Lannion, March 2005. Report available on : http://www.meteorologie.eu.org/safnwc/publis.htm

- Garrett T. J., B. C. Navarro, C. H. Twohy, E. J. Jensen, D. H. Baumgardner, P. T. Bui, H. Gerber, R. L. Herman, A. J. Hemysfield, P. Lawson, P. Minnis, L. Nguyen, M. Poellot, S. K. Pope, F. P. J. Valero and E. M. Weinstock 2005: Evolution of a Florida Cirrus Anvil. *J. Atmos. Sci.* **62** 2352-2372.

- Goldfarb L., P. Keckhut, M. L. Chanin and A. Hauchecorne 2001: Cirrus climatological results from lidar measurements at OHP (44N, 6E). *Geophys. Res. Let.* **28** 1687-1690.

- Haeffelin M., L. Barthes, O. Bock, C. Boitel, S. Bony, D. Bouniol, H. Chepfer, M. Chiriaco, J. Cuesta, J. Delanoe, P. Drobinski, J. L. Dufresne, C. Flamant, M. Grall, A. Hodzic, F. Hourdin, F. Lapouge, Y. Lemaitre, A. Mathieu, Y. Morille, C. Naud, V. Noel, W. O'Hirok, J. Pelon, C. Pietras, A. Protat, B. Romand, G. Scialom, and R. Vautard 2005: SIRTA, a ground-based atmospheric observatory for cloud and aerosol research. *Annales Geophysicae* **23** 253-275.

- Heymsfield A. J. and L. M. Miloshevich 2003: Parameterizations for the Cross-Sectional Area and Extinction of Cirrus and Stratiform Ice Cloud Particles. *J. Atmos. Sci.* **60** 936-956.

- Hoinka K. P. 1998: Statistics of the Global Tropopause Pressure. *Mon. Wea. Rev.* **126** 3303-3325.

- IPCC: Intergovernmental Panel on Climate Change 2001: The Scientific Basis, edited by: Houghton, J. T., Ding, Y., Griggs, D. J., Noguer, M., vand der Linden, P. J., Dai, X., Maskell, K., and Johnson, C. A., Cambridge University Press, Cambridge, UK, 881 pp., 2001.

- Korolev A. , G. A. Isaac , S. G. Cober , J. W. Strapp and  J. Hallett 2003: Microphysical characterization of mixed-phase clouds , *Q. J. R. Meteorol. Soc*. **129**, pp. 39–65

- Mace G. G., E. E. Clothiaux, T. P. Ackerman, 2001: The Composite Characteristics of Cirrus Clouds: Bulk Properties Revealed by One Year of Continuous Cloud Radar Data. *J. Clim.* **14** 2185-2203.

- Matthias V., Freudenthaler, V., Amodeo, A., Balin, I., Baslis, D., B osenberg, J., Chaikovsky, A., Chourdakis, G., Comeron, A., Delaval, A., de Tomasi, F., Eixmann, R.,

Hagard, A. Konguem, L., Kreipl, S., Matthey, R., Rizi, V., Rodrigues, J. A., Wandinger, U., and Wang, X. 2004: Aerosol lidar intercomparison in the framework of the EARLINET project: 1. Instruments. *Appl. Opt.* **43** 961–976.

- McFarquhar G. M., A. J. Heymsfield, J. Spinhirne, and B. Hart 2000: Thin and Subvisual Tropopause Tropical Cirrus: Observations and Radiative Impacts. *J. Atmos. Sci.* **57** (12) 1841-1853.

- Minnis P., J. Kirk Ayers, R. Palikonda and D. Phan 2004: Contrails, Cirrus Trends, and Climate. *J. Climate* **17**, 1671-1685.

- Morille Y., M. Haeffelin, P. Drobinski and J. Pelon 2006: STRAT: An automated Algorithm to Retrieve the Vertical Structure of the Atmosphere from Single Channel Lidar Data. *J. Atmos. Ocean. Tech, in press.*

- Naud, N., M. Haeffelin, P. Muller, Y. Morille, A. Delaval, Assessment of MISR and MODIS cloud top heights through comparison with a back-scattering lidar at SIRTA. *Geophys. Res. Lett.* **31**, L04114, doi:10.1029/2003GL018976, 2004.

- Nastrom G. D., J. L. Green, M. R. Peterson, and K. S. Gage, 1989: Tropopause Folding and the Variability of the Tropopause Height as Seen by the Flatland VHF Radar. *J. Appl. Met.* **28** 1271-1281.

- Nee J. B., C.N. Len, W. N. Chen, and C. I. Lin, 1998: Lidar observations of the cirrus cloud in the tropopause at Chung-Li (25°N, 121°E). *J. Atmos. Sci.* **55** 2249-2257.

- Noel V., H. Chepfer, M. Haeffelin, Y. Morille 2006: Classification of ice crystal shapes in midlatitude ice clouds from three years of lidar observations over the SIRTA observatory. *J. of Appl. Met.* **63** (11) 2978-2991.

- Peter Th. et al., 2003 : Ultrathin Tropical Tropopause Clouds (UTTCs): I. Cloud morphology and occurrence, *Atmos. Chem. Phys.* **3**, 1083–1091

- Platt C. M. R. 1973 : Lidar and Radiometric Observations of Cirrus Clouds. *J. Atmos. Sci.* **30** 1191-1204.

- Platt C. M. R., S. A. Young, P. J. Manson, G. R. Patterson, S. C. Mardsen and R. T. Austin, 1998: The Optical Properties of Equatorial Cirrus from Observations in the ARM Pilot Radiation Observation Experiment. *J. Atmos. Sci.* **55** 1977-1996.

- Prabhakara C., R. S. Fraser, G. Dalu, M.-L. C. Wu and R. J. Curran, 1988: Thin cirrus clouds: Seasonal distribution over oceans deduced from Nimbus-4 IRIS. *J. Atmos. Sci.* **55** 1977-1996.

- Protat, A., M. Haeffelin, A. Armstrong, Y. Morille, J. Pelon, J. Delanoë, and D. Bouniol, 2006: Impact of conditional sampling and instrumental limitations on the statistics of cloud properties derived from cloud radar and lidar at SIRTA. *Geophys. Res. Let.* **33**, L11805, doi:10.1029/2005GL025340.

- Rossow W. B. and L. C. Garder 1993: Cloud detection using satellite measurements of infrared and visible radiances for ISCCP. *J. Climate* **6** 2341-2369.

- Sassen K. , M. K. Griffin, and G. C. Dodd, 1989: Optical scattering and microphysical properties of subvisible cirrus clouds, and climatic implications. *J. Appl. Meteor.* **28**, 91–98.

- Sassen K. 1991: The polarization lidar technique for cloud research: A review and current assessment. *Bull. Amer. Meteor. Soc.* **72** 1848-1866.

- Sassen K. 1992: Evidence for liquid-phase cirrus cloud formation from volcanic aerosols: Climatic implications. *Science* **257** 516-519.

- Sassen K., G. G. Mace, J. Hallett, and M. R. Poellot, 1998: Corona-producing ice clouds: A case study of a cold mid-latitude cirrus layer. *Appl. Opt.* **37** 1477-1485.

- Schmidt T., G. Beyerle, S. Heise, J. Wickert and M. Rothacher 2006: A climatology of multiple tropopauses derived from GPS radio occultations with CHAMP and SAC-C. *Geophys. Res. Let.* **33** L04808, doi:10.1029/2005GL024600.

- Sprenger M., et al. 2003: Tropopause folds and cross-tropopause exchange: A global investigation based upon ECMWF analyses for the time period March 2000 to February 2001, *J. Geophys. Res.*, **108**(D12), 8518, doi:10.1029/2002JD002587.

- Stephens G. L., D. G. Vane, R. J. Boain, G. G. Mace, K. Sassen, Z. Wang, A. J. Illingworth, E. J. O'Connor, W. B. Rossow, S. L. Durden, S. D. Miller, R. T. Austin, A. Benedetti, C. Mitrescu, 2002: The CloudSat Mission and the A-Train. *B. A. M. S.* **83**, 1771-1790, doi:10.1175/BAMS-83-12-1771.

- Stordal F., G. Myrhe, E. J. G. Stordal, W. B. Rossow, D. S. Lee, D. W. Arlander and T. Svendby 2005: Is there a trend in cirrus cloud cover due to aircraft traffic? *Atmos. Chem. Phys.* **5** 2155-2162.

- Wang Z. and K. Sassen 2001: Cloud type and macrophysical property retrieval using multiple remote sensors. *J. Appl. Met.* **40** 1665-1682.

- Winker D. M., J. Pelon, and M. P. McCormick, 2003 : The CALIPSO mission: Spaceborne lidar for observation of aerosols and clouds, *Proc. SPIE Int. Soc. Opt. Eng.*, 4893, 1–11.

- WMO 1957: Definition of the tropopause, *WMO Bull.* **6** 136.

- Zerefos C. S., K. Eleftheratos, D. S. Balis, P. Zanis, G. Tselioudis and C. Meleti 2003: Evidence of impact of aviation on cirrus cloud formation, *Atmos. Chem. Phys.* **3** 1633-1644.

*Tables*

| Cirrus Category | Frequency in observed clouds (sum=100%) |
|---|---|
| Tropospheric Cirrus | 83% |
| Ceiling Cirrus | 10% |
| Cross-Tropopause Cirrus | 5% |
| Inter-Tropopause Cirrus | 2% |

Table 1: Frequencies of occurrence in observed clouds for the 4 cirrus categories.

**Figure captions**

Figure 1: Annual frequencies of multiple tropopauses over the SIRTA observatory, averaged on the 2002-2006 period.

Figure 2: Annual evolution of the mean tropopause altitudes (thin line) and the mean inter-tropopause thickness (thick line) during the 2002-2006 period.

Figure 3: a) Distribution of distance between the first tropopause and cross-tropopause cirrus cloud tops, b) ratio of the cloud top to first tropopause distance on the distance between first and second tropopauses.

Figure 4: Distribution of distance between cloud base and the first tropopause for cross- and inter-tropopause cirrus categories.

Figure 5: a) Backscattering coefficients observed by the LNA lidar on March 17 2005 as a function of time and altitude, using a logarithmic color scale. b) Temperature profile from radiosoundings on March 17 2005. On both figures the first two tropopauses are indicated using dashed lines.

Figure 6: Distribution of cloud Geometrical Thickness, for observations in the May to December period (dark) and December to April (light).

Figure 7: Cumulative density functions of optical depth frequencies for 4 categories of cirrus, together and individually.

Figure 8: Density Contour plots of the frequency of cirrus clouds as a function of cloud geometrical thickness (x-axis) and cloud optical depth (y-axis). Dots are relevant to clouds with base above the first Tropopause.

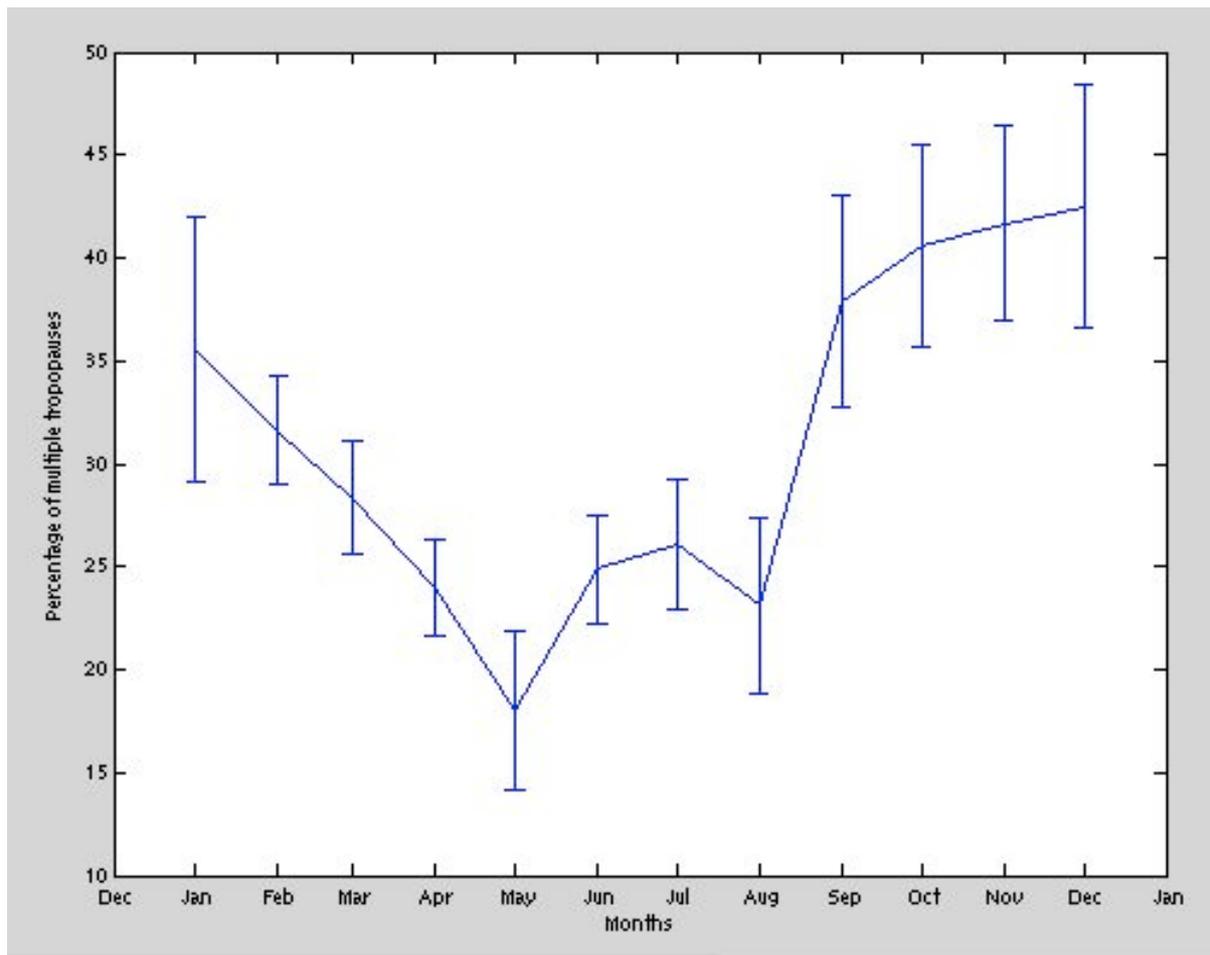

Figure 1: Annual frequencies of multiple tropopauses over the SIRTA observatory, averaged on the 2002-2006 period.

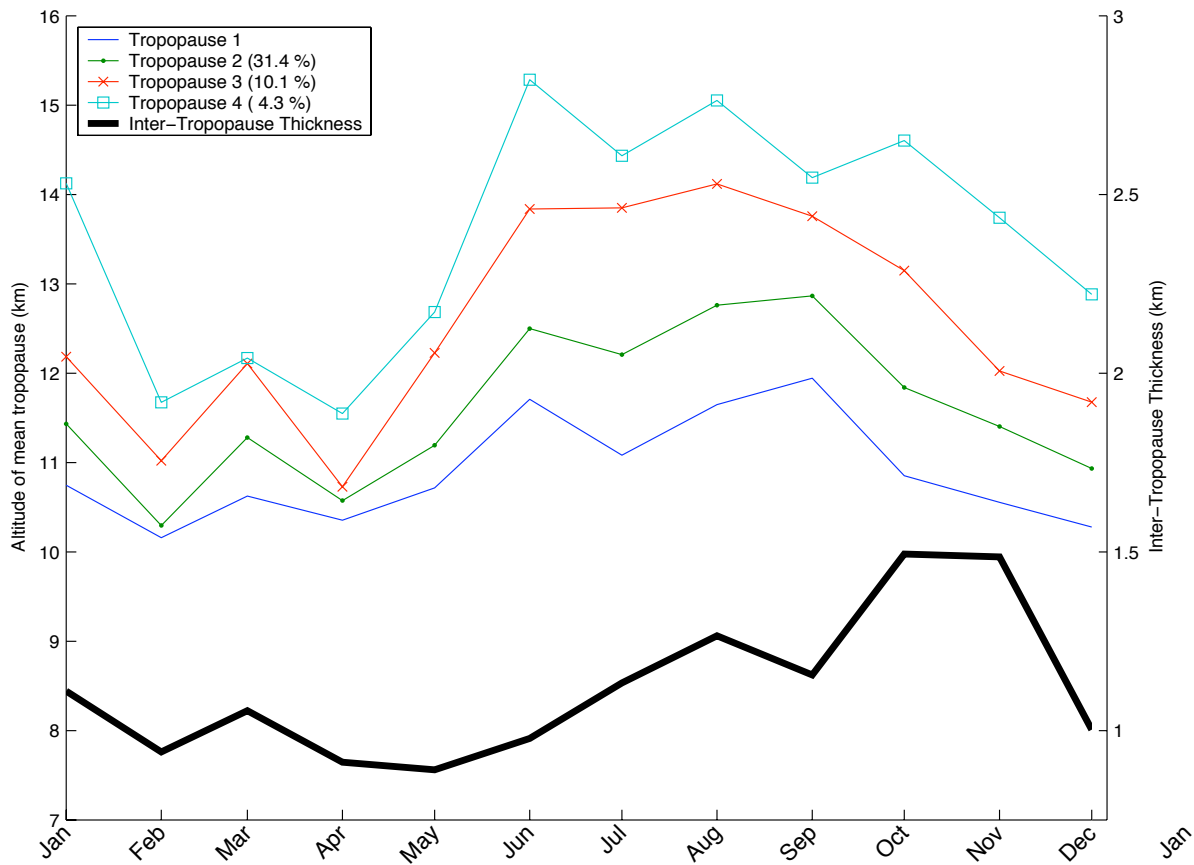

Figure 2: Annual evolution of the mean tropopause altitudes (thin line) and the mean inter-tropopause thickness (thick line) during the 2002-2006 period.

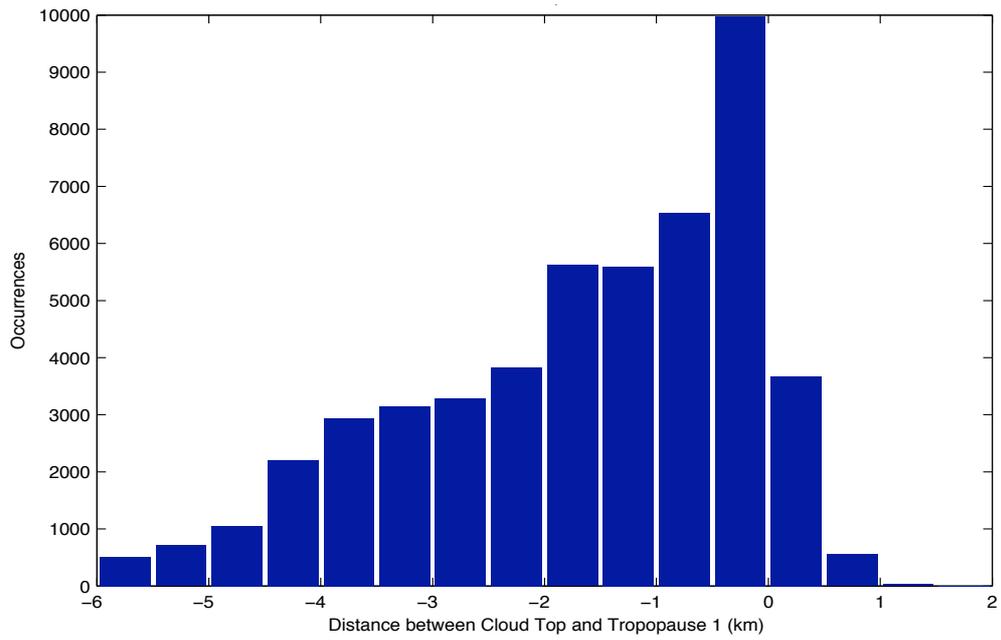

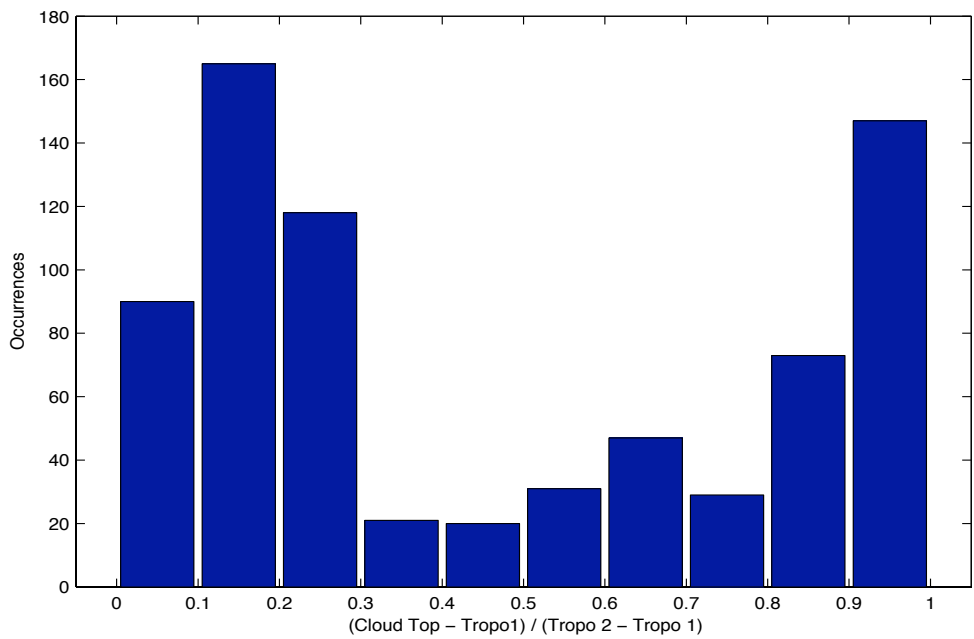

Figure 3: a) Distribution of distance between the first tropopause and cross-tropopause cirrus cloud tops, b) ratio of distance between cloud top and first tropopause, to the distance between the first and second tropopauses.

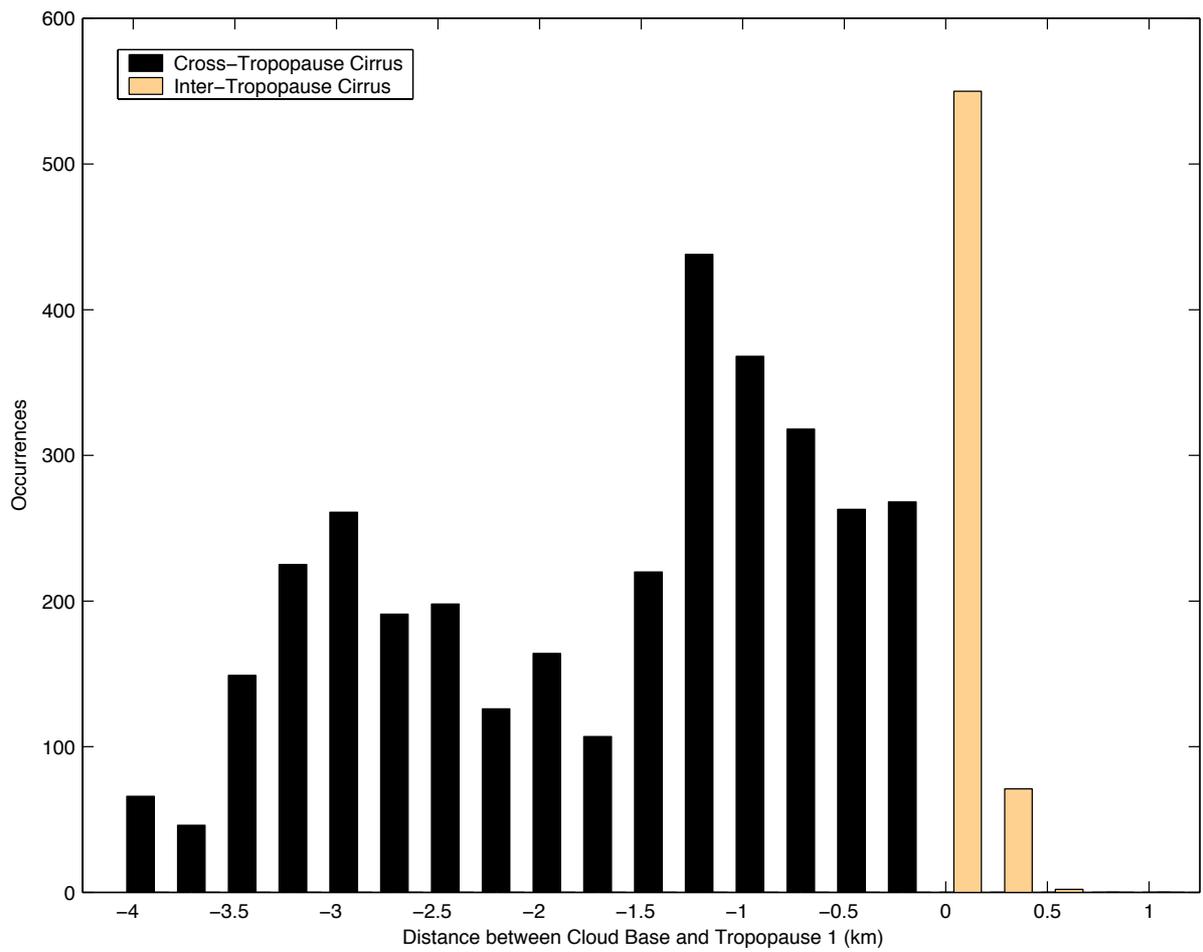

Figure 4: Distribution of distance between cloud base and the first tropopause for cross- and inter-tropopause cirrus categories.

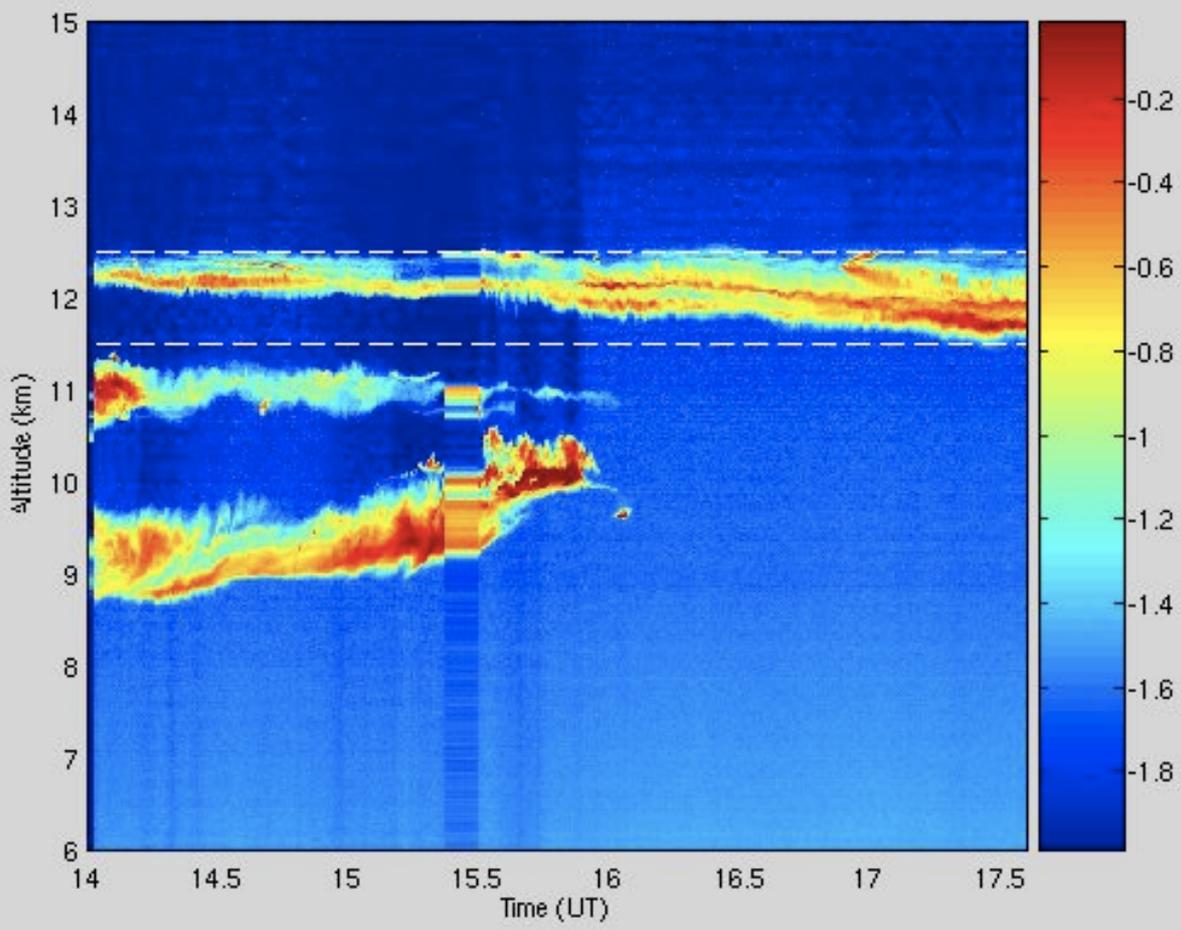

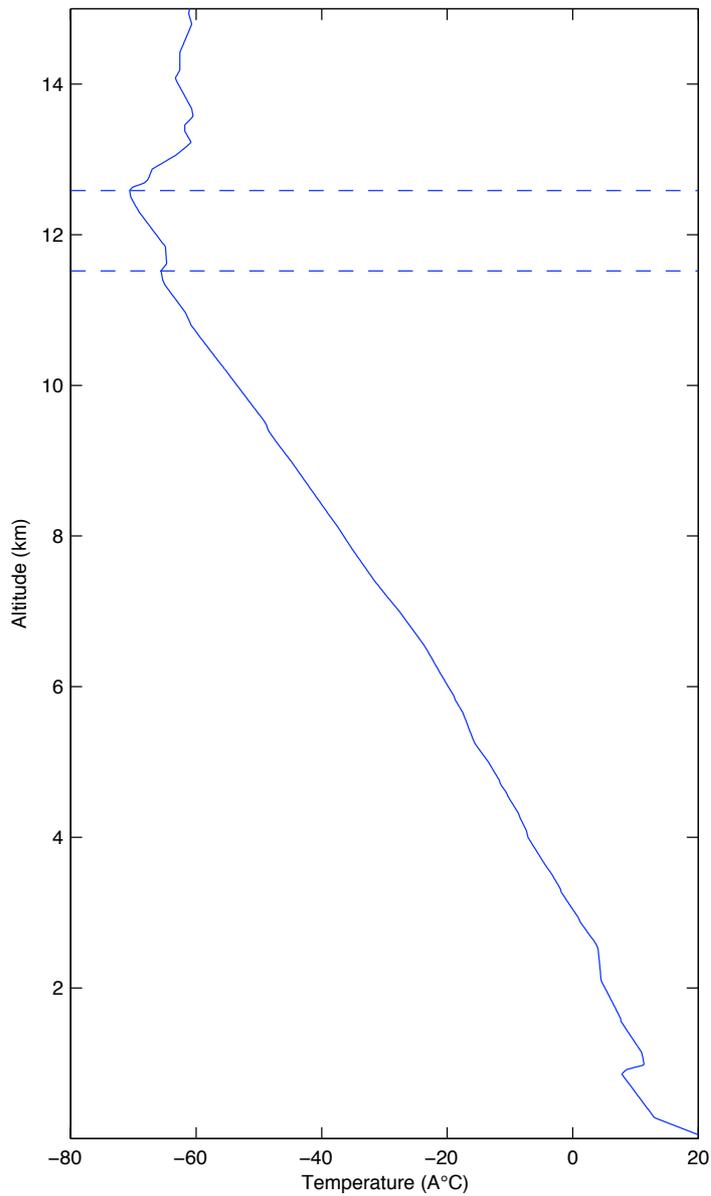

Figure 5: a) Backscattering coefficients observed by the LNA lidar on March 17 2005 as a function of time and altitude, using a logarithmic color scale. b) Temperature profile from radiosoundings on March 17 2005. On both figures the first two tropopauses are indicated using dashed lines.

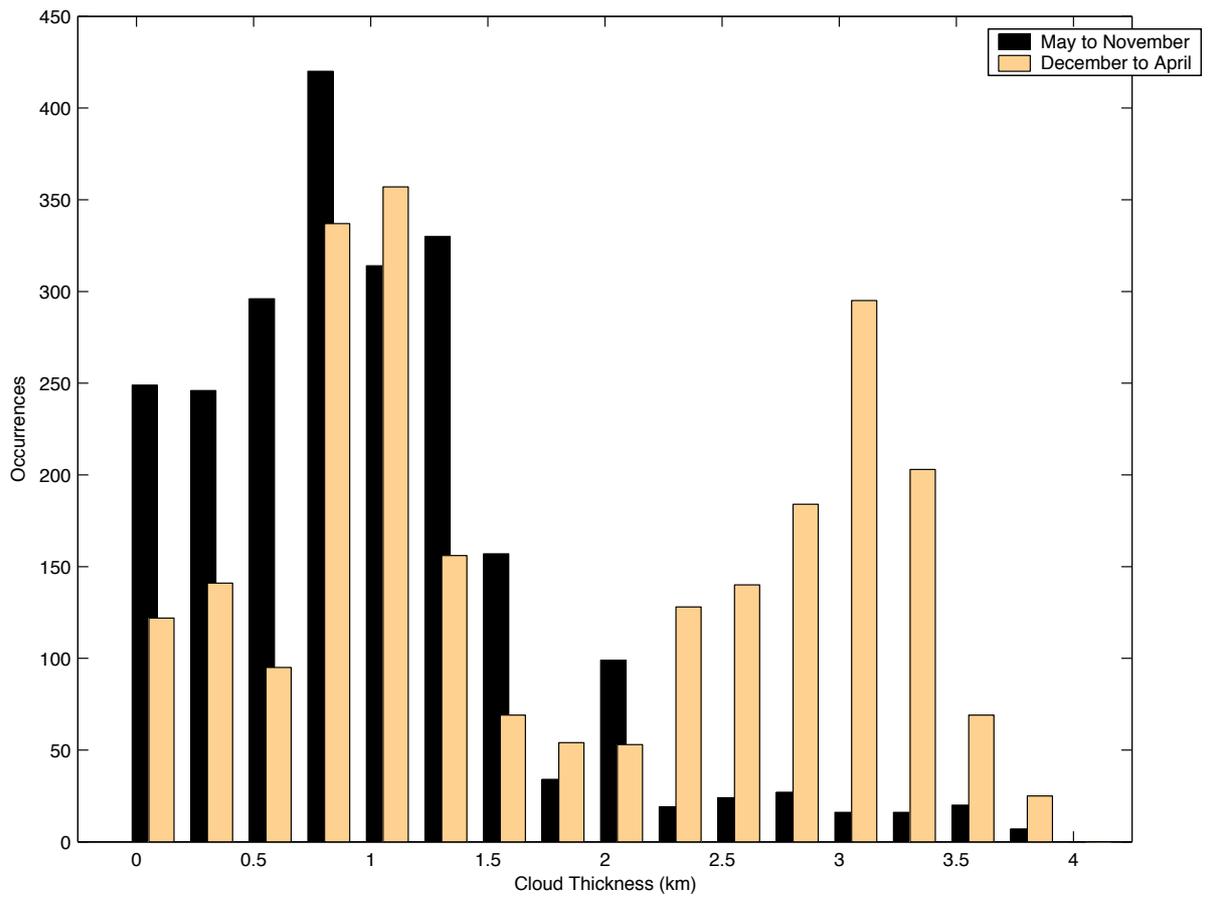

Figure 6: Distribution of cloud Geometrical Thickness, for observations in the May to December period (dark) and December to April (light).

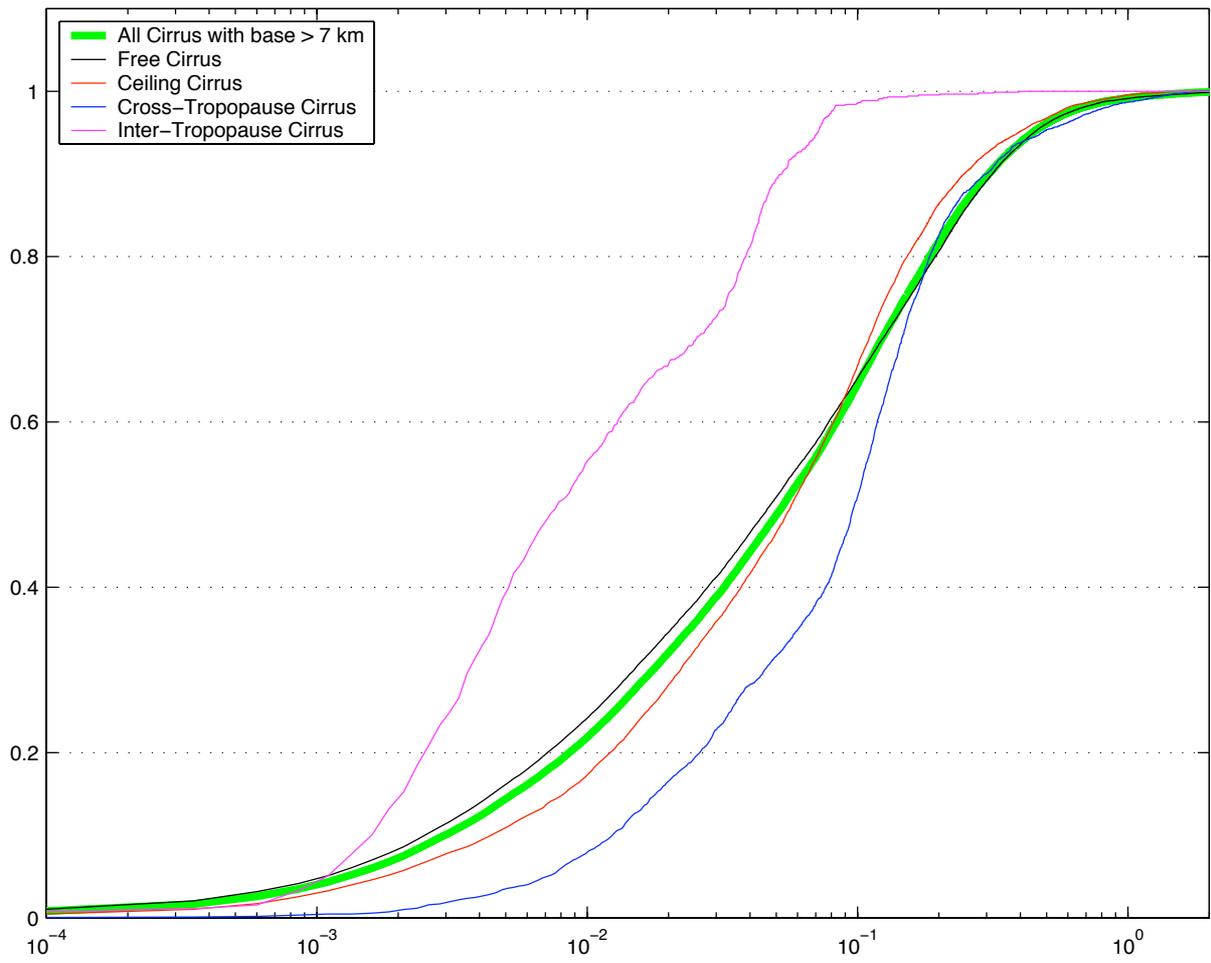

Figure 7: Cumulative density functions of optical depth frequencies for 4 categories of cirrus, together and individually.

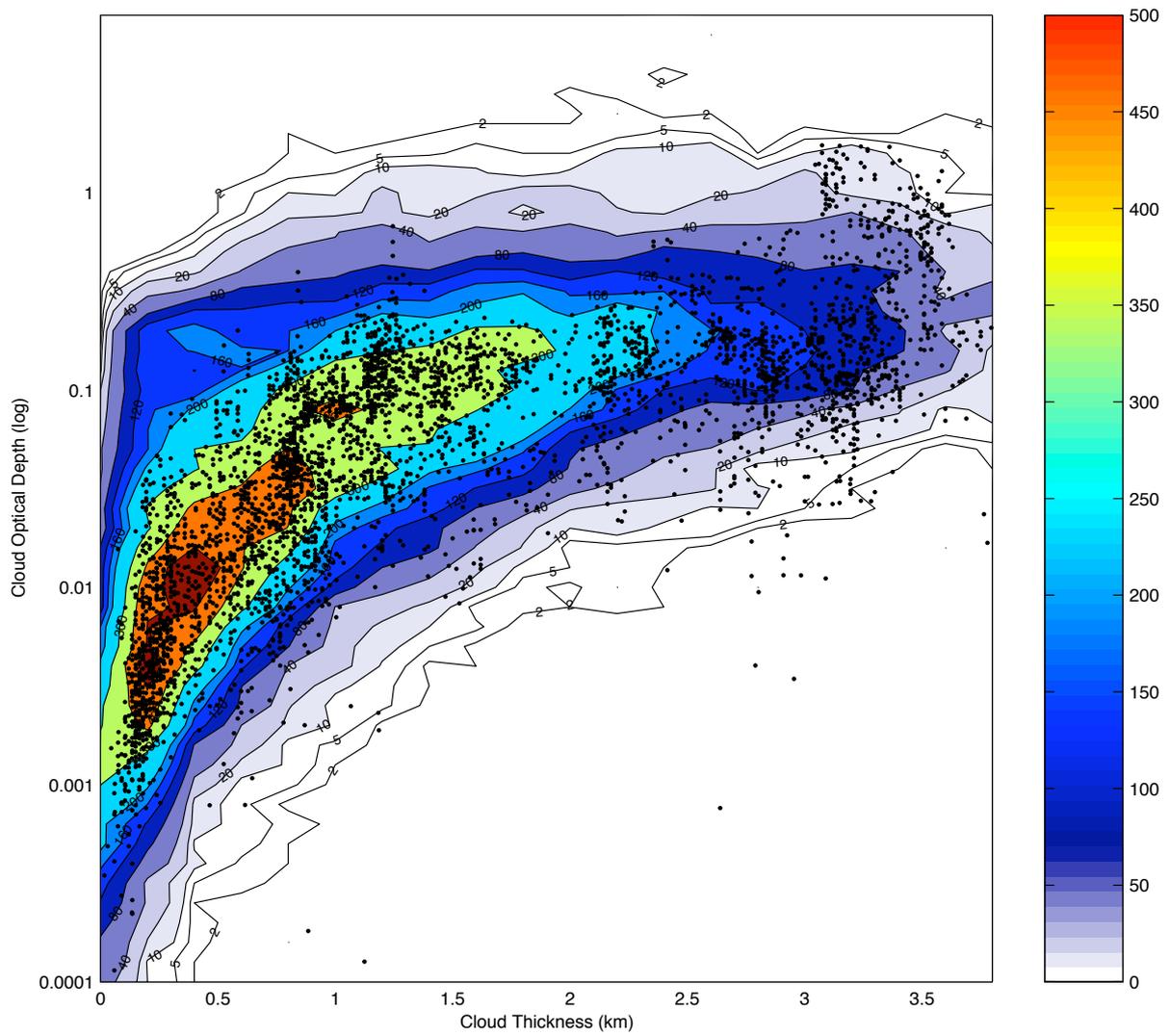

Figure 8: Density Contour plots of the frequency of cirrus clouds as a function of cloud geometrical thickness (x-axis) and cloud optical depth (y-axis). Dots are relevant to clouds with base above the first Tropopause.